\documentclass[10pt]{article} 
\usepackage[preprint]{tmlr}

\usepackage{amsmath,amsfonts,bm}

\def\eqref#1{equation~\ref{#1}}

\def\1{\bm{1}}

\DeclareMathAlphabet{\mathsfit}{\encodingdefault}{\sfdefault}{m}{sl}
\SetMathAlphabet{\mathsfit}{bold}{\encodingdefault}{\sfdefault}{bx}{n}

\usepackage{hyperref}
\usepackage{url}
\usepackage[pdftex]{graphicx}
\usepackage{amsmath}
\usepackage{amsthm}
\newtheorem{theorem}{Theorem}[section]

\usepackage{booktabs}
\usepackage{multirow}
\usepackage{xcolor}
\usepackage{subcaption}
\title{Stochastic Equilibrium Propagation for Spiking Convergent Recurrent Neural Networks}

\author{Jiaqi Lin \email jkl6467@psu.edu \\
\addr School of Electrical Engineering and Computer Science\\
The Pennsylvania State University, University Park, PA 16802, USA
\AND
Yi Jiang  \email ymj5185@psu.edu \\
\addr School of Electrical Engineering and Computer Science\\
The Pennsylvania State University, University Park, PA 16802, USA
\AND
Abhronil Sengupta \email sengupta@psu.edu  \\
\addr School of Electrical Engineering and Computer Science\\
The Pennsylvania State University, University Park, PA 16802, USA}

\def\month{05}  
\def\year{2026} 
\def\openreview{\url{}} 

\begin{document}

\maketitle

\begin{abstract}
Spiking Neural Networks (SNNs) promise energy-efficient, sparse, biologically inspired computation. Training them with Backpropagation Through Time (BPTT) and surrogate gradients achieves strong performance but remains biologically implausible. Equilibrium Propagation (EP) provides a more local and biologically grounded alternative. However, existing EP frameworks for SNNs largely rely on deterministic neurons, which require complex mechanisms to handle spiking discontinuities and do not scale beyond simple benchmarks such as MNIST and Fashion-MNIST. Inspired by the stochastic nature of biological spiking mechanism and recent hardware trends, we propose a stochastic EP framework that integrates probabilistic spiking neurons into the EP paradigm. This formulation smoothens the optimization landscape, stabilizes training, and enables efficient and scalable learning in SNNs. We provide theoretical guarantees showing that the proposed stochastic EP dynamics approximate deterministic EP under mean-field theory, thereby inheriting its underlying theoretical guarantees. The proposed framework narrows the performance gap with both BPTT-trained SNNs and EP-trained non-spiking convergent recurrent neural networks (CRNNs) on CIFAR-10, DVS Gesture, and IMDB datasets, while preserving temporal and spatial locality. Our results highlight stochastic EP as a promising approach for neuromorphic and on-chip learning.
\end{abstract}


\section{Introduction}

\begin{figure*}
  \centering
\includegraphics[width=\textwidth]{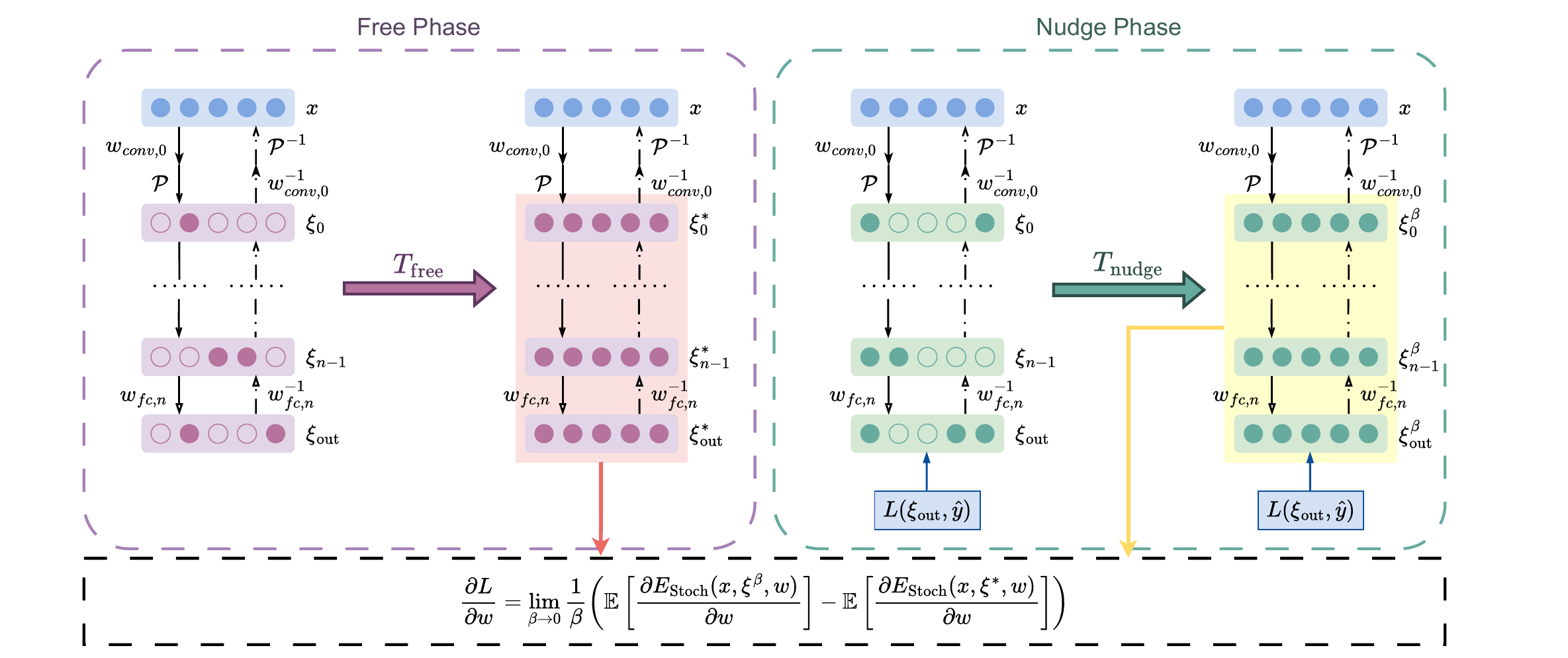}
\caption{Equilibrium Propagation (EP) optimizes neural networks through two phases. With inputs $x$ clamped, the network state $\xi$ relaxes to a fixed point $\xi^{*}$ after $T_{\mathrm{free}}$ time steps. A weak teaching signal then nudges only the output units $\xi_{\mathrm{out}}$ toward their target labels $\hat{y}$ by adding a small perturbation term to the dynamics, producing a nearby fixed point $\xi^{\beta}$ for a small $\beta>0$ after $T_{\mathrm{nudge}}$ time steps. Each synapse updates based on the contrast between its equilibrium states across the two phases. Circles with outlines represent unsaturated neurons, while filled circles denote saturated ones. Purple and green correspond to the free and nudged phases, respectively.}
  \label{fig:ep_overview}
\end{figure*}

Spiking Neural Networks (SNNs) communicate through event-driven spikes, offering substantial energy savings from sparse activity and closer alignment with biological behaviors \citep{mead1990neuromorphic, merolla2014million, benjamin2014neurogrid}. However, most state-of-the-art training methods still rely on Backpropagation Through Time (BPTT) with surrogate gradients \citep{neftci2019surrogate, he2020comparing, yin2021accurate, bal2024spikingbert, fang2021deep, zhou2023spikformer}. \textcolor{black}{Despite their strong performance, BPTT-based approaches remain biologically implausible and computationally inefficient \cite{werbos1990backpropagation, williams1995gradient, decolle_kaiser2020synaptic, eprop_bellec2020solution, lu2022neuroevolution}. These limitations include the weight transport problem \cite{lillicrap2016random, nokland2016direct, frenkel2021learning}, which requires reusing feedforward weights for backward error propagation, global error propagation \cite{baldi2017learning, decolle_kaiser2020synaptic, crick1989recent, lillicrap2020backpropagation}, where gradients depend on computations across layers, and separate circuits for forward inference and backward derivative computation \citep{crick1989recent}.} This motivates Equilibrium Propagation (EP) \citep{scellier2017equilibrium} as a biologically grounded alternative (Figure~\ref{fig:ep_overview}). EP operates a unified circuit that relaxes neuronal states to low-energy fixed points across two phases. Small nudges at output layer implicitly propagate error information backward through feedback connections in convergent recurrent neural networks (CRNNs) with bidirectional connectivity. Synaptic weights are updated from the contrast between pre-perturbation and post-perturbation states in a Spike Timing Dependent Plasticity (STDP)-like, spatially and temporally local manner. These properties make EP particularly attractive for neuromorphic and on-chip implementations \citep{bi1998synaptic, martin2021eqspike, ji2020towards}.

However, extending EP to the spiking domain introduces unique challenges. The binary activations and discontinuous spiking dynamics of SNNs pose challenges for deploying EP, which fundamentally relies on smooth membrane potentials for stable optimization. To address this, predictive coding mechanisms and step-size scheduling have been employed to align discrete SNN dynamics with their continuous counterparts at equilibrium \citep{o2019training, lin2024scaling}. A relaxed, lightweight Leaky Integrate-and-Fire (LIF) neuron formulation has also been explored within EP \citep{martin2021eqspike}. Yet, scalability remains limited, and accuracy degrades as layers deepen, with evaluations largely restricted to simple visual benchmarks. This highlights an open challenge to extend EP to richer architectures and modern datasets while maintaining stability without introducing additional complexity.

Furthermore, current energy-based SNN frameworks \citep{martin2021eqspike, o2019training, lin2024scaling} primarily employ deterministic neurons \citep{lapicque1907louis} with rate-based coding schemes. However, these formulations complicate optimization due to discontinuities arising from spike generation, refractory membrane potentials, and membrane potential resetting mechanisms \citep{izhikevich2003simple}. In contrast, stochastic spiking modulation \citep{jiang2024stochastic}, inspired by the biological observation that neurons fire probabilistically \citep{maass2015spike}, remains largely underexplored in energy-based SNNs. Within the context of energy-based models, particularly under EP, optimization efficiency heavily depends on the smoothness of membrane potentials. Stochastic SNNs address this challenge by preserving smooth membrane dynamics and alleviating discontinuities compared to deterministic neurons. Moreover, optimizing over continuously differentiable likelihoods of the spike generation process \citep{jiang2024stochastic, stochasticSNN} makes direct training of stochastic models practical in both BPTT and EP frameworks.

Motivated by these observations, we develop a stochastic EP framework for training spiking networks. We provide a principled bridge between stochastic spiking networks and EP by \emph{deriving an EP-compatible objective via mean-field analysis}. Rather than relying on an ad-hoc stabilization, we \emph{verify the mean-field conditions} and relate the stochastic spiking dynamics to an equivalent rate-model objective, yielding an \emph{energy-consistent, theoretically grounded} learning update. The contributions of this work are:
\begin{enumerate}
\item To the best of our knowledge, this is the first framework that trains stochastic SNNs within the EP paradigm, extending EP beyond deterministic LIF formulations.
\item The proposed framework achieves stable training and scales to deep convolutional spiking CRNNs, addressing the depth and instability challenges observed in prior EP-trained SNNs.
\item We provide theoretical guarantees showing that the proposed framework is equivalent to the deterministic EP formulation under mean-field theory, ensuring convergence and approximating BPTT.
\item The framework achieves state-of-the-art performance among EP-trained SNNs and BPTT-trained SNNs and narrows the gap to EP-trained non-spiking counterparts, while maintaining reduced computational cost.
\end{enumerate}

\section{Related Work}

\subsection{Stochastic SNNs}
While most studies on SNNs rely on deterministic neuron models, substantial evidence indicates that biological neurons fire stochastically \citep{faisal2008noise, maass2015spike}. Inspired by this observation, stochastic SNNs incorporate probabilistic mechanisms into spike generation or synaptic transmission, offering richer representational capacity and improved robustness. Early works established the theoretical link between stochastic spiking dynamics and probabilistic inference. For example, \citet{buesing2011neural} showed that recurrent SNNs implement Markov Chain Monte Carlo sampling, while \citet{pecevski2011probabilistic} demonstrated approximate inference in graphical models. \citet{ma2023exploiting} further explored noisy SNNs that leverage intrinsic neural noise as a computational resource for scalable and reliable computation.

More recently, stochastic SNNs have been extended to large-scale learning tasks. \citet{jiang2024stochastic} employed stochastic LIF neurons with First-To-Spike coding to improve robustness and reduce latency, whereas \citet{yao2025training} utilized a stochastic Spike Response Model with expectation propagation to enable efficient probabilistic training for supervised tasks. Additionally, \citet{bal2024p} proposed a probabilistic spiking state-space model that leverages stochastic spike generation to efficiently capture long-range dependencies.
Meanwhile, emerging device trends in neuromorphic hardware have further fueled interest in stochastic SNNs \citep{sengupta2016probabilistic, yang2020stochastic}. State-compressed SNNs operating in the probability domain (mapped to scaled stochastic devices) achieve accuracy comparable to multi-bit deterministic counterparts \citep{islam2024hardware, islam2023hybrid}, while stochasticity itself enhances fault tolerance \citep{ardakani2021fault} and mitigates overfitting \citep{hinton2012improving}. Collectively, these studies underscore the promise of stochastic spiking neurons in addressing diverse computational tasks. Consequently, coupling the benefits of stochasticity with EP represents a compelling direction for advancing neuromorphic computing and enabling on-chip learning.

\subsection{Local Learning Algorithms}

\textcolor{black}{Several local learning approaches have been proposed to reduce the reliance on biologically implausible global error propagation in BPTT \citep{lin2025benchmarking, lin2025toward}. Feedback Alignment (FA) replaces symmetric feedback weights with fixed random matrices \citep{lillicrap2016random}, Direct Feedback Alignment (DFA) directly propagates output errors to hidden layers \citep{nokland2016direct, eprop_bellec2020solution}, and Local Error (LE) methods assign layer-wise objectives using local pseudo-targets \citep{decolle_kaiser2020synaptic, frenkel2021learning}.} EP provides another biologically grounded and theoretically equivalent alternative to BPTT \citep{scellier2017equilibrium}. Analyses show that EP closely approximates BPTT \citep{scellier2017equilibrium, scellier2019equivalence}. However, early EP variants suffered from first-order bias under finite nudging, restricting architectures to shallow fully connected networks \citep{scellier2017equilibrium, o2019training, ernoult2019updates, ernoult2020equilibrium}. Three-phase training procedures have been proposed to mitigate this bias \citep{laborieux2021scaling} and reduce the performance gap with BPTT-based baselines on both vision and sequential NLP tasks \citep{lin2025scalable, bal2023equilibrium}. Collectively, these approaches advance biological plausibility and computational efficiency, making them promising candidates for neuromorphic and on-chip learning.

\subsection{EP in Spiking CRNN Architectures}
EP has also been explored in spiking CRNNs. Prior work has primarily relied on deterministic spiking neurons, particularly LIF models \citep{martin2021eqspike, o2019training, lin2024scaling}. A central challenge in these approaches lies in smoothing membrane potentials to maintain stable optimization within the EP framework. For instance, \citet{martin2021eqspike} introduced a low-pass moving average filter for LIF neurons, but this method was restricted to shallow linear layers, where deeper stacking led to significant accuracy degradation. Alternatively, predictive encoding and decoding strategies with step-size scheduling have been proposed to better align discrete-time SNN dynamics with the fixed points of non-spiking CRNNs \citep{o2019training, lin2024scaling}. Despite these efforts, EP-trained SNNs have struggled to scale beyond shallow architectures and simple visual benchmarks, with accuracy degrading as network depth increases. This persistent limitation underscores the need for approaches that stabilize spiking dynamics while preserving EP’s spatial and temporal locality. To the best of our knowledge, the proposed framework is the first to integrate stochastic neurons directly into EP, combining the smooth membrane potential dynamics of stochastic SNNs with the locality advantages of energy-based learning.

\section{Methods}

\subsection{Stochastic Spiking Neuron}

\begin{figure}
  \centering
  \includegraphics[width=0.5\textwidth]{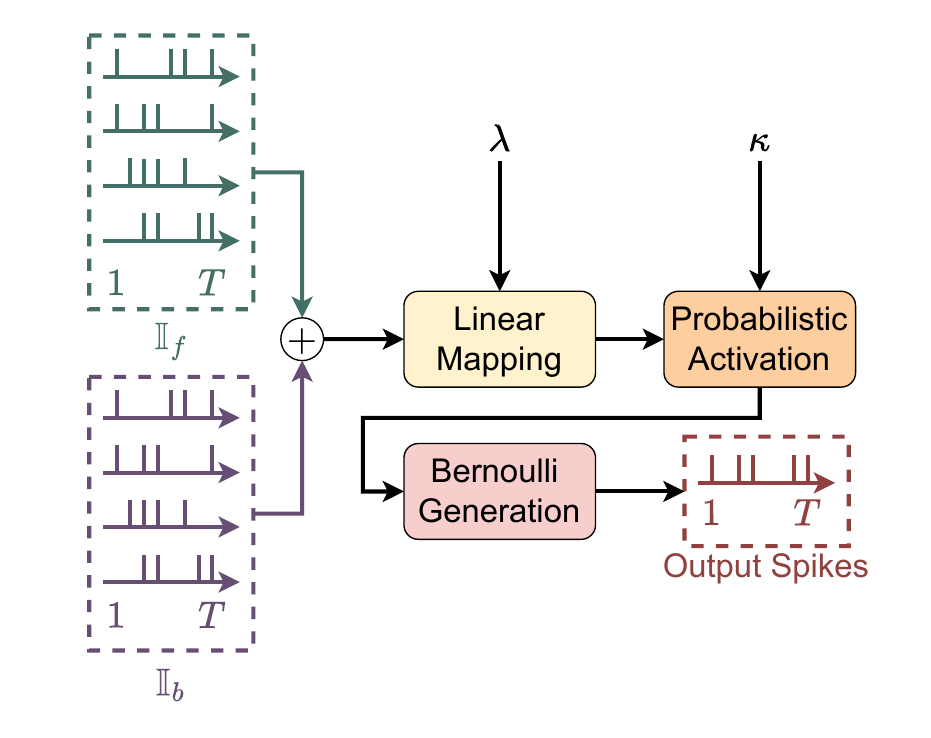}
  \caption{Illustration of stochastic spiking neuron dynamics. The membrane potential integrates weighted inputs from both forward $\mathbb{I}_f$ and backward passes $\mathbb{I}_b$ with decay factor $\lambda$, which is mapped to a firing probability scaled by factor $\kappa$, and generates spikes through Bernoulli sampling.
 }
  \label{fig:stochastic_neurons}
\end{figure}

Inspired by prior work \citep{jiang2024stochastic, stochasticSNN, binaryEP}, we design stochastic spiking neurons (Figure~\ref{fig:stochastic_neurons}) that are compatible with the proposed stochastic energy function. The membrane potential $\xi^t$ is formulated as a linear combination of the previous membrane potential and the input contributions across the temporal dimension (the formula will be derived later in Equation~\ref{eq:nd1}).
Subsequently, a hard sigmoid function is applied to compute the firing probability at time $t$, and spikes $s^t \in \{0,1\}$ are generated through Bernoulli sampling:
\begin{align}
\label{eq:act}
    s^t &= \mathcal{B}(\sigma(\xi^t))\\ 
    \label{eq:sigma}
    \sigma(x) &= \operatorname{clip}(\kappa x, 0, 1),
\end{align}
where $\kappa$ is a tunable scaling factor that controls the firing frequency\textcolor{black}{, and $x$ denotes the generic input argument.  The clipping operation, defined as $\operatorname{clip}(x,0,1)=\min(\max(x,0),1)$, ensures that $\sigma(x)\in[0,1]$, yielding a valid Bernoulli firing probability.} Although spike generation is inherently non-differentiable, this issue is mitigated in our proposed EP framework because gradients are computed from local changes in the membrane potential $\xi^t$, which is a continuous variable. In contrast to deterministic LIF neurons, stochastic spiking neurons naturally exhibit smoother dynamics (Section~\ref{sec:neuron_changes}), as they avoid hard resets and refractory periods that cause abrupt state changes and complicate optimization within the EP framework.

\subsection{Stochastic Energy Function and Neuron Dynamics}

We now formalize the proposed stochastic EP framework as an underlying energy-based model. It builds on a continuous-valued Hopfield network with symmetric recurrent connections, characterized by the Hopfield energy function \citep{hopfield1984neurons, scellier2017equilibrium}. Extending this formulation to layered architectures \citep{binaryEP, ernoult2019updates, scellier2019equivalence} yields the following energy function \textcolor{black}{$E(x,\xi^t,w)\in\mathbb{R}$} (we omit the bias term and input layer for simplicity):
\begin{equation}
\label{eq:base_engr}
E(x,\xi^t,w) = \tfrac{1}{2} \sum_{i=0}^{N_{\mathrm{t}}-1} {\|\xi^t_i\|}^2 - \sum_{i=0}^{N_{\mathrm{t}}-2} \rho(\xi^t_{i})^\mathsf{T}  w_i \rho(\xi^t_{i+1})
\end{equation}
Here, \textcolor{black}{$\xi_i^t \in \mathbb{R}^{n_i}$} denotes the neuron states at layer $i$, \textcolor{black}{where $n_i$ is the number of neurons in that layer}, \textcolor{black}{$w_i\in \mathbb{R}^{n_i \times n_{i+1}}$} represents the synaptic weight matrix between layers $i$ and $i+1$, $\rho(\cdot)$ is the activation function, and $N_{\mathrm{t}}$ denotes the total number of layers.
The corresponding dynamics are given by
\begin{equation}
\frac{\partial \xi^{t}}{\partial t} = - \frac{\partial E(x,\xi^t,w)}{\partial \xi}
\end{equation}
To incorporate stochastic spiking behavior into the computational circuit of EP, which provides high sparsity and significantly reduces computational cost (Section~\ref{sec:cost}), we extend Equation~\ref{eq:base_engr} into a stochastic version of the energy function by replacing the deterministic activation function with a stochastic spike generation process:
\begin{equation}
\label{eq:sto_engr}
E_{\rm Stoch}(x,\xi^t,w) =\tfrac{1}{2} \sum_{i=0}^{N_{\mathrm{t}}-1} {\|\xi^t_i\|}^2
 - \sum_{i=0}^{N_{\mathrm{t}}-2} \mathcal{B}(\sigma(\xi^t_{i}))^\mathsf{T}  w_i \mathcal{B}(\sigma(\xi^t_{i+1}))
\end{equation}
Here, $\mathcal{B}(\cdot)$ denotes Bernoulli sampling and $\sigma(\cdot)$ is a probabilistic activation function. To address the non-differentiability of Bernoulli sampling, the straight-through (ST) estimator \citep{ste1,ste2} for $\mathcal{B}^\prime(\cdot)$ is employed, thereby yielding valid gradients for stochastic neurons. Derivation of neuron dynamics based on Equation~\ref{eq:sto_engr} is defined as:
\begin{equation}
\begin{split}
  &  \frac{\partial \mathcal{B}(\sigma(\xi^t))}{\partial \xi^t} =  \mathcal{B}^\prime(\sigma(\xi^t)) \sigma^\prime(\xi^t)\approx 1 \cdot \sigma^\prime(\xi^t)\\
   & \frac{\partial \xi_i^{t}}{\partial t}= \sigma^\prime(\xi^t_i)\Big(w_{i}\mathcal{B}(\sigma(\xi^t_{i+1})) + w_{i-1}^\mathsf{T}\mathcal{B}(\sigma(\xi^t_{i-1}))\Big) - \xi^t_i
\end{split}
\end{equation}
Based on the Euler method with step size $\lambda$ (interpretable as the decay rate of the membrane potential), the discretized neuron dynamics can be expressed as:
\begin{equation}
\xi^{t+1}_i = (1-\lambda)\xi^{t}_i  +\lambda\sigma^\prime(\xi^{t}_i) \Big(w_{i}\mathcal{B}(\sigma(\xi^{t}_{i+1})) + w_{i-1}^\mathsf{T}\mathcal{B}(\sigma(\xi^{t}_{i-1}))\Big)
\label{eq:nd1}
\end{equation}
With the straight-through estimator approximation, Equation~\ref{eq:nd1} preserves the stochastic spiking generation process in inter-neuron communication through $\mathcal{B}(\sigma(\cdot))$, while the gradient computation of the activation depends solely on the continuous function $\sigma(\cdot)$. The convergence of the proposed neuron dynamics is demonstrated in Section~\ref{sec:neuron_changes}.

\subsection{Equilibrium Propagation}
The training procedure of EP is adapted to perform gradient descent in spiking CRNNs on a loss function $L(\xi^t_{\rm out}, \hat{y})$ defined between the target $\hat{y}$ and the output activations $\xi^t_{\rm out}$ (Figure~\ref{fig:ep_overview}) \citep{ernoult2019updates, o2019training, lin2024scaling, martin2021eqspike}.
In the first (free) phase, a static input $x$ is presented, and the network evolves over $T_{\rm free}$ time steps until it converges to a stable state $\xi^*$ that minimizes the stochastic energy function $E_{\rm Stoch}(x,\xi^t,w)$. In the second (nudge) phase, a nudging term $\beta \frac{\partial L(\xi^t_{\rm out}, \hat{y})}{\partial \xi}$ with $\beta$ denoting the nudging factor, is applied to the output layer $\xi^t_{\rm out}$. The resulting neuron dynamics are given by
\begin{equation}
\label{eq:nudge_neuron}
\frac{\partial \xi^{t}}{\partial t} = -\frac{\partial E_{\rm Stoch}(x,\xi^{t},w)}{\partial \xi} - \beta \frac{\partial L(\xi^t_{\rm out}, \hat{y})}{\partial \xi}
\end{equation}         
This nudging term shifts the saturated states of the network toward another stable state $\xi^{\beta}$, which lies closer to the true label after $T_{\rm nudge}$ time steps. \textcolor{black}{The nudging factor $\beta$ is selected empirically to provide sufficient separation between the free and nudged states while maintaining stable convergence.}

Then the weight gradients can be computed by 
\begin{equation}
\label{eq:noisy} 
\frac{\partial L}{\partial w} = \lim_{\beta \to 0}\frac{1}{\beta}\left(  \frac{\partial E_{\rm Stoch}(x,\xi^{\beta},w)}{\partial w} - \frac{\partial E_{\rm Stoch}(x,\xi^*,w)}{\partial w}\right)
\end{equation}
However, accurate estimation of weight gradients \citep{scellier2017equilibrium, ernoult2019updates} depends on stable neuron states, whereas Equation~\ref{eq:noisy} yields a noisy optimization landscape due to fluctuations introduced by the stochastic spike generation process. In extreme cases, $\sum_{i=0}^{N_{\mathrm{t}}-2} \mathcal{B}(\sigma(\xi^t_{i}))^\mathsf{T}  w_i \mathcal{B}(\sigma(\xi^t_{i+1}))$ may vanish when no spikes are generated at time $t$. To address this issue, we establish the equivalence between the proposed stochastic energy function and its deterministic counterpart under mean-field theory (see Appendix~\ref{app:expect_proof} for the formal proof), as stated in the following theorem.

\begin{theorem}
\label{thm:stoeng}
Suppose the energy function (Equation~\ref{eq:base_engr} and Equation~\ref{eq:sto_engr}) has symmetric weights. The mean firing rate satisfies:
\begin{equation}
    \mathbb{E}[\mathcal{B}(\sigma(\xi^t_i))] = \sigma(\xi^t_i)
\end{equation}
Under a mean-field independence assumption for sufficiently large networks \citep{sompolinsky1988chaos,buice2010systematic}, we assume the units are independent across neuronal indices ($j\not=k$).
\begin{equation}
\mathbb{E}[\mathcal{B}(\sigma(\xi^t_j)) \mathcal{B}(\sigma(\xi^t_k))] = \sigma(\xi^t_j)\sigma(\xi^t_k)
\end{equation}
Then, given that neuron states $\xi$ are deterministic, and $\sigma(\cdot) = \rho(\cdot)$, the expected stochastic energy (Equation~\ref{eq:sto_exp_engr}) equals the deterministic energy function. 
\begin{equation}
\mathbb{E}[E_{\rm Stoch}(x,\xi^t,w)] = E(x,\xi^t,w)
\end{equation}
where
\begin{equation}
\label{eq:sto_exp_engr}
\mathbb{E}[E_{\rm Stoch}(x,\xi^t,w)] = \tfrac{1}{2} \sum_{i=0}^{N_{\mathrm{t}}-1} {\|\xi^t_i\|}^2
- \sum_{i=0}^{N_{\mathrm{t}}-2}\sigma(\xi^t_{i})^\mathsf{T}  w_i \sigma(\xi^t_{i+1})
\end{equation}
\end{theorem}
Leveraging Theorem~\ref{thm:stoeng} mitigates the instability caused by directly computing weight gradients from $E_{\rm Stoch}$. It is worth mentioning here that the independence of neuronal activities is a common assumption in neural network models \citep{la2021mean}. We make two key observations:
\begin{enumerate}
\item The expected stochastic energy coincides with the deterministic energy under mean-field theory, effectively constructing a deterministic rate model.
\item The neuron dynamics (Equation~\ref{eq:nd1} and Equation~\ref{eq:nudge_neuron}) resemble a moving average of firing rates, where the saturated states of stochastic neurons in the free and nudged phases represent their mean firing rates. This correspondence yields stable states for gradient computation.
\end{enumerate}
Based on these observations, we estimate the weight gradients in stochastic EP using the expected energy function, which approximates the gradients of a rate-based deterministic EP and therefore preserves its theoretical foundations, as stated in the following theorem.
\begin{theorem}
\label{thm:weight}
The gradient of the objective function $L$ with respect to $w$ can be estimated from the divergence of the two stable states:
\begin{equation}
\label{eq:weight}
\frac{\partial L}{\partial w} = \lim_{\beta \to 0}\frac{1}{\beta} \Bigg( \mathbb{E} \left[\frac{\partial E_{\rm Stoch}(x,\xi^{\beta},w)}{\partial w}\right] - 
 \mathbb{E} \left[\frac{\partial E_{\rm Stoch}(x,\xi^*,w)}{\partial w}\right]\Bigg)
\end{equation}
\end{theorem}
We refer the reader to \citet{scellier2019equivalence} for the theoretical justification of Theorem~\ref{thm:weight} in deterministic settings, and to \citet{ernoult2019updates} and \citet{laborieux2021scaling} for explicit formulations of the input functions in convolutional and linear layers.

\subsection{Augmenting the Number of Output Neurons}
\label{sec:outputneuron}
\citet{binaryEP} argues that, in EP frameworks with binary activations, effective backward error transmission to upstream layers requires a sufficient number of output neurons whose activation function changes satisfy:
\begin{equation}
\Delta \xi= |\xi^\beta - \xi^{*}| > \frac{1}{2}
\end{equation}
To achieve this, we follow prior work by increasing the error signal through output-layer augmentation, where each prediction neuron is replaced by $N_{\rm perclass}$ neurons per class, inflating the output layer from $N_{\rm classes}$ to $N_{\rm classes} \times N_{\rm perclass}$. \textcolor{black}{$N_{\rm{perclass}}$ is selected empirically to ensure a sufficiently strong nudging signal and avoid vanishing weight updates, while avoiding unnecessary output-layer expansion.}

\section{Experiments}
\label{sec:experiments}

\begin{figure*}
  \centering
  \includegraphics[width=\textwidth]{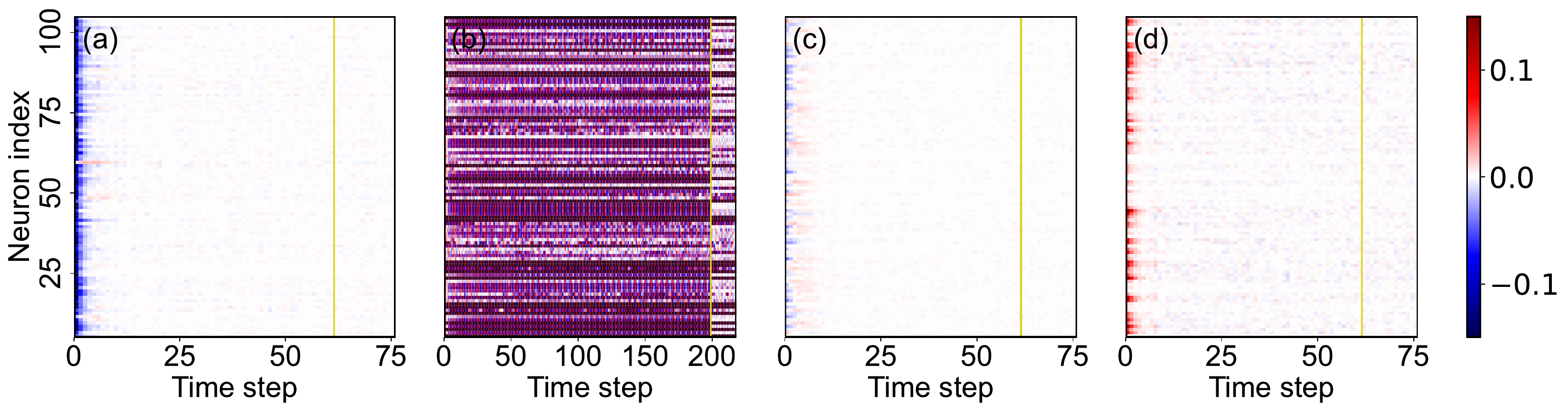}
  \caption{Comparison of membrane potential stability across different spiking neuron models. Heatmaps show changes in membrane potential over time for a network with one hidden layer of 512 neurons trained on 100 random MNIST samples. (a) The proposed stochastic model exhibits stable dynamics, rapidly converging to a smooth equilibrium. (b) A deterministic LIF model with a low-pass filter \citep{martin2021eqspike} shows pronounced instability. (c) A deterministic LIF model stabilized using predictive coding and step-size scheduling \citep{o2019training, lin2024scaling} achieves convergence. (d) A deterministic LIF model with only step-size scheduling, simplified from \citep{o2019training, lin2024scaling}, exhibits fluctuations. The yellow vertical dashed line indicates the transition from the free phase to the nudged phase.
 }
  \label{fig:stab_mem}
\end{figure*}

\begin{table}
  \caption{Testing accuracy (\%) of spiking and non-spiking CRNNs trained using the EP framework on MNIST, CIFAR-10, DVS Gesture, and IMDB datasets. Results are reported as the mean over five independent runs, with standard deviation shown in parentheses. Here, \#FC denotes the number of hidden fully connected layers (excluding output layers) and \#C the number of convolutional layers. \#HA refers to the number of Hopfield attention layers. \emph{Bin.} (Y/N) indicates whether the model employs binary activations, \emph{Optim.} denotes the optimization method, and \emph{Local} (Y/N) specifies whether the optimization approach is a local learning method.}
  \label{tab:perf}
  \centering
  \begin{subtable}{\linewidth}
      \centering
      \begin{tabular*}{\linewidth}{@{\extracolsep{\fill}}cccccc}
        \toprule
        \textbf{Model} & \textbf{Bin.} &  \textbf{Optim.} & \textbf{Local}  & \textbf{Accuracy}\\
        \midrule
        1FC\citep{o2019training} & Y& EP & Y & 97.63\\
        1FC\citep{martin2021eqspike} & Y& EP & Y & 97.59\\
        2FC\citep{binaryEP} & Y& EP & Y & 96.97\\
        2C\citep{ernoult2019updates} & N& EP & Y  & 98.98\\
        2C1FC\citep{lin2024scaling} & Y & EP & Y & 99.03\\
        2C1FC\citep{lin2024scaling} & Y& BPTT & N  & 99.14\\
        2C1FC\citep{lin2024scaling} & N & BPTT & N  & 99.01\\
        1FC (Ours) & Y & EP & Y  & 97.46\\
        2FC (Ours) & Y & EP & Y  & 98.05\\
        2C (Ours) & Y & EP & Y  & 98.99\\
        \bottomrule
      \end{tabular*}
      \caption{Evaluated on the MNIST dataset.}
    \end{subtable}
  
    \begin{subtable}[t]{\linewidth}
      \centering
      \begin{tabular*}{\linewidth}{@{\extracolsep{\fill}}cccccc}
        \toprule
        \textbf{Model} & \textbf{Bin.} &  \textbf{Optim.} & \textbf{Local}  & \textbf{Accuracy}\\
        \midrule
        VGG-5\citep{laborieux2021scaling} &N& EP &Y  & 87.55\\
        VGG-5\citep{binaryEP} &N& EP &Y & 84.34\\
        VGG-5\citep{laborieux2022holomorphic}&N& BPTT  &N  & 88.30\\
        5C \citep{massa2020efficient}&Y& BPTT  & N  & 78.92\\
        ReActNet \citep{chen2021bnn} & Y & BPTT  & N  & 77.60\\
        5C \citep{fa} &Y& FA  & N  &  71.48 (0.42) \\
        5C \citep{xiao2021training} &Y& ID  & Y  &  76.25 (0.39)\\
        5C (Ours) &Y& EP & Y  & 81.33 (0.18)\\
        5CP (Ours) &Y& EP & Y & 81.73 (0.19)\\
        \bottomrule
      \end{tabular*}
      \caption{Evaluated on the CIFAR-10 dataset.}
    \end{subtable}
    
    \begin{subtable}[t]{\linewidth}
      \centering
      \begin{tabular*}{\linewidth}{@{\extracolsep{\fill}}cccccc}
        \toprule
        \textbf{Model} & \textbf{Bin.} &  \textbf{Optim.} & \textbf{Local} & \textbf{Accuracy}\\
        \midrule
        2C1FC\citep{viale2021carsnn} &Y & BPTT  &N  & 78.70\\
        2C1FC\citep{stewart2020online} &Y & SOEL  & Y  & 64.70\\
        2C1FC\citep{stewart2020online} &Y & SLAYER\citep{shrestha2018slayer}  & N & 83.50\\
        3C (Ours) &Y  & EP  &Y & 89.92 (0.74)\\
        \bottomrule
      \end{tabular*}
      \caption{Evaluated on the DVS Gesture dataset.}
    \end{subtable}
    \begin{subtable}{\linewidth}
      \centering
      \begin{tabular*}{\linewidth}{@{\extracolsep{\fill}}cccccc}
        \toprule
        \textbf{Model} & \textbf{Bin.} &  \textbf{Optim.} & \textbf{Local} & \textbf{Accuracy}\\
        \midrule
        1HA2FC\citep{bal2023equilibrium} &N & EP & Y & 88.9 \\
        1HA1FC (Ours) & Y & EP & Y & 80.99 (0.30)\\
        \bottomrule
      \end{tabular*}
      \caption{Evaluated on the IMDB dataset.}
    \end{subtable}
\end{table}

\subsection{Stability of Stochastic Spiking CRNN}
\label{sec:neuron_changes}
In this section, we conduct a toy experiment using a single hidden layer of 512 neurons trained on the MNIST dataset to analyze membrane potential stability across different spiking neuron models. A single-layer configuration is chosen due to the scalability limitations of the deterministic baseline \citep{martin2021eqspike}. The objective is to show that stochastic spiking neurons trained with the proposed stochastic EP framework converge to stable states and exhibit higher stability than deterministic LIF neurons. All models are trained to reach at least 75\% accuracy to ensure convergence, and membrane potentials are averaged over 100 randomly selected MNIST samples.

Figure~\ref{fig:stab_mem} compares the stability of the proposed stochastic model against deterministic baselines. The heatmaps track the temporal evolution of membrane potentials as the network transitions from the free phase to the nudged phase. Our stochastic model consistently converges to an equilibrium with minimal potential variance across neurons, whereas deterministic methods encounter notable stability issues. The LIF model with a low-pass filter \citep{martin2021eqspike} shows strong instability. Incorporating predictive coding and step-size scheduling \citep{o2019training, lin2024scaling} improves stability in deterministic models but increases computational cost (Section~\ref{sec:cost}). Removing predictive coding reveals that step-size scheduling alone cannot maintain stable membrane dynamics. These findings demonstrate that the stochastic formulation provides a more efficient and stable solution for EP training in SNNs.

\subsection{Dataset and Network Topology \label{sec:setup}}
The proposed framework is evaluated on standard datasets, including MNIST, CIFAR-10, DVS Gesture, and IMDB datasets (details in Appendix~\ref{app:datasets}). The architectural configurations for each dataset are provided in Appendix~\ref{app:topology}. Model-specific hyperparameters were optimized individually for optimal performance and summarized in Appendix~\ref{app:hyper}. All experiments were implemented in Python using the PyTorch framework and executed on an NVIDIA RTX A5000 GPU with 24 GB of memory. The resulting performance is reported in Table~\ref{tab:perf}.

{\color{black}
\subsection{Correlation of Neuron Firing Events}

\label{sec:firing_correlation}
In this section, we empirically assess the independence assumption in Theorem~\ref{thm:stoeng} by measuring the pairwise Pearson correlation between binary neuron firing events. For a given time step $t$, the Pearson correlation $\mathcal{P}_{ij}^t$ between neuron $i$ and neuron $j$ over a batch of $N_s$ samples is defined as:
\begin{equation}
\mathcal{P}_{ij}^t = \frac{\sum_{k=1}^{N_s} (s_{i,k}^t - \bar{s}_i^t)(s_{j,k}^t - \bar{s}_j^t)}{\sqrt{\sum_{k=1}^{N_s} (s_{i,k}^t - \bar{s}_i^t)^2} \sqrt{\sum_{k=1}^{N_s} (s_{j,k}^t - \bar{s}_j^t)^2}}
\end{equation}
Here, $s_{i,k}^t$ represents the binary firing state of neuron $i$ for sample $k$, and $\bar{s}_i^t$ denotes the mean firing activity of neuron $i$ averaged across all $N_s$ samples.
At each time step, we compute this correlation coefficient for all possible neuron pairs and report the mean absolute pairwise correlation. Figure~\ref{fig:firing_correlation} presents these results for the 2C and 5C architectures on the MNIST and CIFAR-10 datasets, respectively. The consistently low correlations observed across time steps and network layers indicate that the neuron firing events are largely uncorrelated, providing empirical support for the mean-field independence assumption.

\begin{figure}
  \centering
  \includegraphics[width=0.9\textwidth]{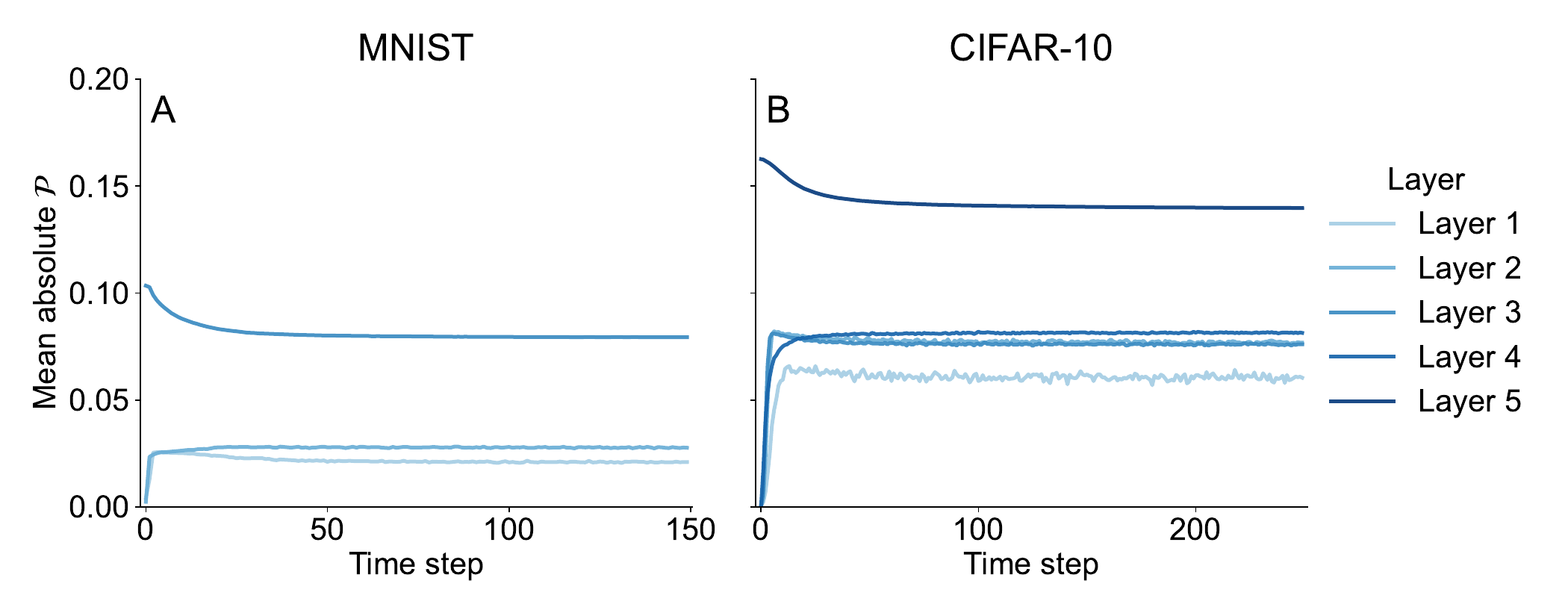}
  \caption{\textcolor{black}{Mean absolute Pearson correlation of binary firing events over time. At each time step, pairwise correlations are computed across 1,000 randomly selected samples and averaged across all neuron pairs. Results are presented for (A) the 2C architecture on the MNIST dataset and (B) the 5C architecture on the CIFAR-10 dataset.}}
  \label{fig:firing_correlation}
\end{figure}
}

\subsection{Performance}
\label{sec:perf}
The proposed stochastic EP framework substantially narrows the performance gap between non-spiking models and their BPTT-trained counterparts (Table~\ref{tab:perf}). To the best of our knowledge, this work presents the first evaluation of EP-trained SNNs on complex datasets such as CIFAR-10, DVS Gesture (see Appendix~\ref{app:eval_dvs} for additional details), and IMDB datasets. \textbf{Notably, the framework achieves comparable performance compared to iso-architecture SNNs trained with BPTT.} Although its accuracy on CIFAR-10 and IMDB remains approximately 3\% to 8\% lower than that of full-precision models, it offers significantly reduced computational and memory costs (see Section~\ref{sec:mem_consump} and Section~\ref{sec:cost}). \textcolor{black}{We attribute this performance gap between the proposed framework and full-precision models to the information loss and limited representational capacity associated with binary spike activations, which have been shown to contribute to accuracy degradation \citep{guo2024ternary}.} We further evaluate an alternative stochastic mechanism for SNNs based on a Poisson escape-rate process \citep{amarasingham2006spike, buxo2020poisson, ding2025neuromorphic}, which yields performance comparable to Bernoulli spike generation (denoted \textsc{5CP} in Table~\ref{tab:perf}), demonstrating the generalizability of the proposed formulation. 
Compared to other local learning approaches, including feedback alignment (FA) \citep{fa} and implicit differentiation (ID) \citep{xiao2021training}, our framework achieves superior performance on CIFAR-10. (Note that the results for FA and ID are produced using our re-implementation of the original methods in order to ensure a consistent architecture across all comparisons.) Overall, these results highlight the effectiveness, generalizability, and scalability of the proposed stochastic EP framework.
\subsection{Memory Consumption}
\label{sec:mem_consump}
The GPU memory consumption of spiking CRNNs trained with EP and BPTT is reported in Table~\ref{tab:mem}. All experiments were conducted with a batch size of 128 using the architectures described in Section~\ref{sec:setup}. The number of time steps for each model was set according to the optimal hyperparameters yielding the best performance, as detailed in Appendix~\ref{app:hyper}. The results demonstrate that the EP framework offers a more memory efficient training scheme for deep SNNs. EP achieves substantial memory savings while maintaining performance comparable to BPTT-trained counterparts. Notably, these memory savings remain significant even when the dimensionality of the output layers is inflated.

\begin{table*}
\centering
\caption{GPU memory consumption (in MB) of CRNNs trained using the EP framework on the MNIST, CIFAR-10, DVS Gesture, and IMDB datasets. \#FC represents the number of hidden linear layers, and \#C denotes the number of convolutional layers. \#HA refers to the number of Hopfield attention layers. The batch size is set to 128 for the MNIST, CIFAR-10, and DVS Gesture datasets, and to 64 for the IMDB dataset. The notation $N\times$ indicates that BPTT consumes $N$ times more GPU memory than EP.}
\label{tab:mem}
\begin{tabular}{c|c|c|c|c|c|c}
\toprule\hline
\textbf{Dataset} & \multicolumn{3}{c|}{MNIST} & CIFAR-10 & DVS Gesture  & IMDB\\
\hline
\textbf{Model}& 1FC & 2FC & 2C & 5C & 3C & 1HA1FC\\
\hline
  EP & 141  & 147  & 281 & 7885  & 4617 & 12267\\
\hline
\multirow{2}{*}{BPTT} & 179 &  195 & 677 & 42938 & 10381 & 18189\\
&(1.2$\times$)& (1.3$\times$)  &(2.4$\times$) & (5.4$\times$) & (2.2$\times$) & (1.5$\times$)\\
\hline
\bottomrule
\end{tabular}
\end{table*}

\section{Ablation Studies}

\subsection{Computational Cost in Stochastic EP Spiking Models}
\label{sec:cost}
We quantify computational complexity using the number of Multiply–Accumulate Operations (MACs) and accumulation operations (ACs). In this work, we estimate MACs and ACs using the same formulation for both forward and backward computational paths.

\paragraph{Computational cost compared to full-precision models.}
We estimate inference energy efficiency on neuromorphic hardware by comparing the number of spiking computations with the operation count of an iso-architecture full-precision model trained with EP (Table~\ref{tab:mem}). Following standard cost formulations \citep{lu2020exploring, howard2017mobilenets, sandler2018mobilenetv2}, the number of operations for the $i$-th layer in an full-precision network (unidirectional) is defined as 
\begin{equation}
    \mathrm{MAC}^{\rm full-precision}_i = 
    \begin{cases}
    C_{\text{in}} K_H K_W C_{\text{out}} O_H O_W 
    &\text{if  } 1\leq i\leq N_{\rm c}\\
    I_F O_F &\text{if  } N_{\rm c} < i \leq N_{\rm t} \\
    \end{cases}
\end{equation} 
Here, $C_{\text{in}}$ and $C_{\text{out}}$ are the input and output channels, $K_H \times K_W$ is the kernel size, $O_H \times O_W$ are the output dimensions, and $I_F$, $O_F$ are the input and output features. $N_{\rm c}$ and $N_{\rm t}$ denote the number of convolutional and total layers, respectively. The total number of operations (unidirectional) in the full-precision model is therefore $\mathrm{MAC}^{\rm full-precision}_{\rm tot} = \sum_{i=1}^{N_{\rm t}} \mathrm{MAC}^{\rm full-precision}_i$.
Unlike full-precision networks, SNNs operate on sparse, event-driven signals, where computation occurs only when spikes are generated. For a layer with an average firing rate $IFR_i$, the number of synaptic operations (unidirectional) in the proposed stochastic framework is $\mathrm{AC}^{\rm SNN}_{\rm tot} = \sum_{i=1}^{N_{\rm t}} IFR_i \times \mathrm{MAC}^{\rm full-precision}_i$. 

We estimate the total energy consumption on the MNIST dataset using the hyperparameters in Appendix~\ref{app:hyper} for the 2FC and 2C architectures. For CRNNs, the computation of $\mathrm{MAC}^{\rm full-precision}_{\rm tot}$ and $\mathrm{AC}^{\rm SNN}_{\rm tot}$ is performed in both forward and backward directions. To evaluate the worst-case scenario, we assume the SNN architectures employ inflated output neurons (700 neurons at output layer), while the full-precision models use the number of classes as output neurons (10 neurons at output layer). The average firing rates $IFR_i$ are [0.21, 0.19, 0.12] and [0.27, 0.19, 0.10] for the 2FC and 2C architectures, respectively (averaged over 128 samples). Using hardware energy costs of 0.9 pJ for ACs and 4.6 pJ for MACs \citep{energyaccmac}, total energy consumption is computed by multiplying the energy per operation by the corresponding operation count. Under these settings, the proposed framework achieves 19× and 23× reductions in energy consumption relative to full-precision models.

\paragraph{Computational cost compared to deterministic spiking model.}
We analyze the computational efficiency of the proposed method against deterministic SNNs trained with EP. Prior works on scalable spiking CRNNs \citep{o2019training, lin2024scaling} integrate predictive coding into neuron dynamics within the EP framework, formulated as
\begin{equation}
\label{eq:snn_nd}
\begin{split}
\boxed{Dec^t} &= (1-\alpha)\colorbox{red}{$\cdot$}Dec^{t-1} \colorbox{yellow}{$+$} \alpha\colorbox{red}{$\cdot$}\mathbb{I}\\
\xi^t &= (1-\lambda)\xi^{t-1} + \lambda\rho'(\xi^{t-1})Dec^t\\
\boxed{Enc^t} &= {\frac{1}{\alpha}\colorbox{red}{$\cdot$}\big[\rho(\xi^t)} \colorbox{yellow}{$-$} (1-\alpha)\colorbox{red}{$\cdot$}\rho(\xi^{t-1})\big]\\
s^t &= \delta(V^{t-1} + Enc^t > \mathcal{V}_{th})\\
\boxed{V^t} &= V^{t-1} \colorbox{yellow}{$+$} Enc^t \colorbox{yellow}{$-$} s^t
\end{split}
\end{equation}
where $\mathbb{I}$ is the weighted inputs, and $\delta(\cdot)$ governs deterministic spike generation based on the membrane potential $V^t$ and threshold $\mathcal{V}_{th}$. The predictive decoder estimates the current signal $Dec^t$ from previous samples with predictive factor $\alpha$, and the predictive encoder $Enc^t$ quantizes temporal changes in neuron states into binary signals. Assuming a total of $I$ neurons and equivalent spike generation cost as in the stochastic framework (Equation~\ref{eq:act}), predictive coding adds $4I$ multiplications (red in Equation~\ref{eq:snn_nd})  and $4I$ additions (yellow in Equation~\ref{eq:snn_nd}) per time step to update $Dec^t$ and $Enc^t$. Each layer further requires $3I$ additional memory units to store intermediate variables $Dec^t$, $Enc^t$, and $V^t$ (boxed in Equation~\ref{eq:snn_nd}). These overheads highlight the efficiency of our stochastic formulation, which achieves stable dynamics without predictive coding.

\subsection{Effect of Increasing Number of Output Neurons}

The inflation factor $N_{\rm perclass}$ scales to strengthen error signals. Insufficient learning signals are observed in models with binary activations \citep{binaryEP}, including our stochastic spiking network, where neuron activations can be zeroed out during the backpropagation of error signals. $N_{\rm perclass}$ amplifies the error signal $\frac{\partial L(\xi^t_{\rm out}, \hat{y})}{\partial \xi}$ (Appendix~\ref{app:output_neuron}) to ensure it is sufficiently potent to perturb the network state from its free-phase equilibrium $\xi^*$ towards the nudge-phase equilibrium $\xi^\beta$ at nudged phase. An insufficient nudge yields $\xi^* \approx \xi^\beta$, causing weight updates to vanish. However, merely increasing $N_{\rm perclass}$ does not guarantee better performance. In our experiments, it is carefully tuned to prevent vanishing gradient rather than to optimize performance directly.

\section{Discussion}
The proposed stochastic EP framework shows that incorporating stochastic spiking neurons into Equilibrium Propagation yields stable learning dynamics and competitive performance on standard vision benchmarks. Under mean-field theory, it can be viewed as optimizing an energy landscape over neuronal firing rates, providing theoretical grounding for convergence and equivalence to BPTT. By mitigating discontinuities inherent in deterministic spiking models, stochastic spiking neurons improves training stability and smoothens optimization, narrowing the performance gap between EP and surrogate-gradient BPTT while preserving the efficiency and locality of energy-based learning. These results establish stochasticity as both a biologically plausible and computationally advantageous mechanism for neuromorphic and on-chip learning.
Although the framework scales to deeper convolutional CRNNs, extending it to large-scale architectures and datasets such as ResNet \citep{he2016deep} on ImageNet \citep{imagenet} remains an open challenge. Subsequent research should also encompass hardware-level deployments to fully assess the neuromorphic potential of stochastic EP.

\section*{Acknowledgments}

This material is based upon work supported in part by the U.S. National Science Foundation under award No. CAREER \#2337646 and CCSS \#2333881.

The authors acknowledge the use of LLMs in the preparation of this manuscript. Specifically, GPT-5.4 (OpenAI) were used for wording refinement, and identifying potentially relevant prior work. All technical claims, numerical values, cited references, and final text were independently verified by the authors, who take full responsibility for the manuscript.

\bibliography{main}
\bibliographystyle{tmlr}

\clearpage

\appendix

\section{Evaluation on Neuromorphic Datasets}
\label{app:eval_dvs}
Unlike static datasets, samples in the DVS Gesture dataset \citep{amir2017low} are time-varying event sequences, whereas the standard EP framework is formulated for static inputs. Directly applying EP to temporal data leads to non-convergent energy dynamics because the saturated neuron states vary from frame to frame. To address this, we adapt the framework by introducing separate training and evaluation loops. For each temporal step $\tau$ in an input sequence, both the free and nudged phases are executed for $T_{\rm free}$ and $T_{\rm nudge}$ iterations, respectively, allowing the network to reach equilibrium at every $\tau$. Weight updates are then computed using the saturated states $\xi^{\tau,*}$ and $\xi^{\tau,\beta}$ at each time step. The total runtime per sample becomes $T_{\rm total} = \tau_{\rm total} \times (T_{\rm free} + T_{\rm nudge})$, where $\tau_{\rm total}$ denotes the total number of time steps in each input sequence. Final predictions are obtained by summing the output activations $\xi^{\tau,*}_{\rm out}$ across all $\tau$.

Additionally, we propose a warm-up initialization that relates the initial state for frame at time $\tau+1$ to the saturated state for frame at time $\tau$. Concretely, instead of resetting to zero, we initialize as a linear function of the previous equilibrium:
\begin{equation}
\xi^{\tau+1,0} = \mathbf{0} + \gamma \xi^{\tau,*},
\end{equation}
where $\mathbf{0}$ is the zero state and $\gamma \in [0,1]$ controls the degree of temporal state retention. This design is motivated by the following assumptions:
\begin{enumerate}
    \item \textbf{Temporal similarity:} consecutive frames in DVS Gesture exhibit high similarity due to the underlying event-based dynamics.
    \item \textbf{Stability of equilibrium:} if two static input frames are similar, the network dynamics converge to similar equilibrium ( $\xi^{\tau+1,*} \approx \xi^{\tau,*}$ when inputs are similar).
\end{enumerate}

\section{Hopfield Attention in EP Framework}
\label{sec:hopfield_attention}
 While the above procedure enables EP to operate on event streams, many sequential domains, such as language tasks, are more naturally modeled by explicit sequence mechanisms rather than frame-wise equilibration alone. A widely used approach is to incorporate attention-based modules to capture long-range dependencies. In the EP literature, Hopfield attention has been explored in full-precision settings \citep{bal2023equilibrium}. Here, we adapt and integrate Hopfield attention into our proposed EP framework. Specifically, the state transition dynamics are governed by the Hopfield attention module, formulated as:
\begin{equation}
\xi_i^{t+1} =  (1-\lambda)\xi_i^t +\lambda \operatorname{HopAttn}(Q,K)
\end{equation}
The $\mathrm{HopAttn}$ operator yields an attention-based embedding of the input sequence, defined by scaled dot-product attention:
\begin{equation}
    \mathrm{HopAttn}(Q, K) = \mathrm{softmax}\left(\frac{1}{\sqrt{d_k}} Q K^\mathsf{T}\right)K
\end{equation}
Here, $d_k$ denotes the encoding dimension of the key $K$, \textcolor{black}{and $Q$ is the query.} In our experiments, we apply Hopfield attention only at the first layer. Specifically, $Q$ is formed by adding the initial input $x$, projected into the query space via $W_Q$, to the feedback signals from the next layer. The key $K$ consists of the sequence of stored patterns in the Hopfield network, typically corresponding to the network input $x$. Crucially, the Hopfield attention module carries both theoretical and empirical guarantees of converging to a steady state after a single update step \citep{hopfieldatt, bal2023equilibrium}. By leveraging this property, the system is driven toward stable states at every time step.

\section{Datasets}
\label{app:datasets}
We evaluate on three standard vision benchmarks spanning increasing complexity and sensing modalities: MNIST (static grayscale digits), CIFAR-10 (natural RGB images), and DVS Gesture (event-based neuromorphic recordings). These datasets enable a balanced evaluation across both static and spiking regimes. For the DVS Gesture dataset, we use 60 time steps per sample. We additionally evaluate our framework on IMDB as a language benchmark. Word representations for the IMDB dataset are initialized using 300-dimensional GloVe embeddings \citep{pennington2014glove}. These embeddings serve as input to the model, with input sequences truncated or padded to a maximum length of 256 tokens.

\paragraph{MNIST.}
The MNIST dataset contains 60{,}000 training and 10{,}000 test images of handwritten digits \citep{mnist}.

\paragraph{CIFAR-10.}
CIFAR-10 comprises 32$\times$32 color images across ten object categories, posing a more challenging classification task than MNIST \citep{cifar10}.

\paragraph{DVS Gesture.}
The DVS Gesture dataset includes recordings from 29 subjects captured under three distinct lighting conditions, providing a neuromorphic benchmark with event-driven dynamics \citep{amir2017low}.

\paragraph{IMDB.}
The IMDB dataset consists of 50{,}000 highly polar movie reviews split evenly into 25{,}000 training and 25{,}000 test samples, serving as a standard benchmark for binary sentiment classification \citep{maas2011learning}.

\section{Network Topologies}
\label{app:topology}
In this section, we detail the architectural topology used for each dataset.
For MNIST, we evaluate three architectures. Two fully connected models (1FC and 2FC) with one and two hidden layers of 512 neurons each, and a 2C model composed of two convolutional layers with 64 and 128 channels, followed by a linear layer mapping features to the output dimension. Each convolutional layer uses a kernel size of 5, stride 1, and padding 1, followed by a $3\times3$ max pooling operation with stride 3.
For CIFAR-10, we adopt a 5C architecture consisting of four convolutional layers with 64, 128, 256, and 256 channels, and a linear layer mapping features to the output. Each layer employs a kernel size of 5, stride 2, and padding 2, followed by a $2\times2$ max pooling operation with stride 2. In the \textsc{5CP} setting, spike generation follows a Poisson escape-rate process. The probability of emitting a spike at time $t$ is given by
\begin{equation}
    P(s^t = 1 \mid \xi^t) = 1 - \exp\left(-\sigma(\xi^t)\Delta t\right),
\end{equation}
where $\Delta t = 1$ denotes the simulation time step and $\kappa = 1$ is the activation scaling factor.
For the DVS Gesture dataset, we use a 3C model comprising three convolutional layers with 64, 128, and 256 channels, and a linear layer mapping features to the output. Each convolutional layer applies a kernel size of 5, stride 1, and padding 1, followed by a $2\times2$ max pooling operation with stride 2.
For the IMDB dataset, we employ a single attention module at the first layer (following \citep{bal2023equilibrium}), followed by a linear readout layer that maps the extracted features to the output logits.

\section{Hyperparameters}
\label{app:hyper}
This section outlines the hyperparameter configurations utilized in the experiments. Table~\ref{tab:param} summarizes the hyperparameters used for the MNIST, CIFAR-10, DVS Gesture, and IMDB datasets. For MNIST, the network is optimized using stochastic gradient descent (SGD) without momentum. For CIFAR-10, DVS Gesture, and IMDB datasets, the AdamW optimizer \citep{adamw} is employed with a weight decay of 0.001, 0.0001, and 0.01 respectively.

Additionally, conventional EP \citep{scellier2017equilibrium} introduces an estimation bias due to the positive nudging factor $\beta$, leading to inaccurate gradient estimates. \citet{laborieux2021scaling} addressed this issue by introducing a randomized $\beta$ and a three-phase training procedure. The additional phase applies a nudging factor of $-\beta$ to counteract the bias. In this work, we adopt the same three-phase training strategy. Let $\xi^{-\beta}$ denote the steady state of the network at the end of the third phase. Under this formulation, Equation~\ref{eq:weight} becomes

\begin{equation}
\label{eq:3phase}
\frac{\partial L}{\partial w} = \lim_{\beta \to 0}\frac{1}{2\beta} \Bigg( \mathbb{E} \left[\frac{\partial E_{\rm Stoch}(x,\xi^{\beta},w)}{\partial w}\right] - 
\mathbb{E} \left[\frac{\partial E_{\rm Stoch}(x,\xi^{-\beta},w)}{\partial w}\right]\Bigg)
\end{equation}
which preserves both spatial and temporal locality in updating the synaptic weights between connected layers. The use of randomized $\beta$ and three-phase training is summarized in Table~\ref{tab:param}.

\begin{table*}[htbp]
\caption{Hyper-parameters for optimal performance on various datasets.\label{tab:param} LR represents the learning rates and BS represents the batch sizes.}
\begin{center}

\begin{tabular*}{\linewidth}{@{\extracolsep{\fill}}|c|c c c c c c c c c c c|}
\hline
\textbf{Model}&  $\lambda$& $T_{\rm free}$ & $T_{\rm nudge}$ & $\beta$ & $\kappa$ & $N_{\rm perclass}$ & \textbf{LR} & \textbf{BS} & \textbf{Epoch} & \textbf{Bias Issue} & \textbf{$\gamma$}\\
\hline
\multicolumn{12}{|c|}{\textbf{MNIST}} \\
    \hline
     1FC & 0.5 & 60 & 15 & 0.75 & 2 & 10 & 3e-3 & 4 & 100 & Random & -\\
     2FC & 0.5 & 60 & 15 & 0.5 & 2 & 70 & 2e-2 & 64 & 200 & Random  & -\\
     2C & 0.5 & 150 & 50 & 0.5 & 2 & 70 & 5e-4 & 16 & 200 & Random  & -\\
    \hline
\multicolumn{12}{|c|}{\textbf{CIFAR-10}} \\
    \hline
    5C & 0.3 & 250 & 50 & 0.15 & 1 & 50 & 1e-4 & 100 & 200 & 3-Phase & -\\
    \hline
\multicolumn{12}{|c|}{\textbf{DVS Gesture}} \\
    \hline
    3C & 0.25 & 150 & 35 & 0.25 & 1 & 50 & 1e-4 & 64 & 100 & 3-Phase & 0.3\\
    \hline
\multicolumn{12}{|c|}{\textbf{IMDB}} \\
    \hline
    Attention & 0.5 & 150 & 50 & 0.5 & 1 & 100 & 1e-5 & 25 & 200 & 3-Phase & -\\
    \hline
\end{tabular*}
\end{center}
\end{table*}

\section{Scaling Factor, Firing Density, and Sparsity}

\label{sec:sparsity}
Prior to Bernoulli sampling, activations are scaled by a gain $\kappa$. Larger $\kappa$ increases firing density and correspondingly decreases sparsity. This value provides an effective balance between ensuring sufficient spike activity for information propagation and maintaining a high degree of sparsity, which is a key property that enables SNNs to achieve energy-efficient computation \citep{roy2019towards,zenke2018superspike, esser2016convolutional, davies2018loihi, pehle2022brainscales}.

\begin{figure}
  \centering
  \includegraphics[width=0.7\textwidth]{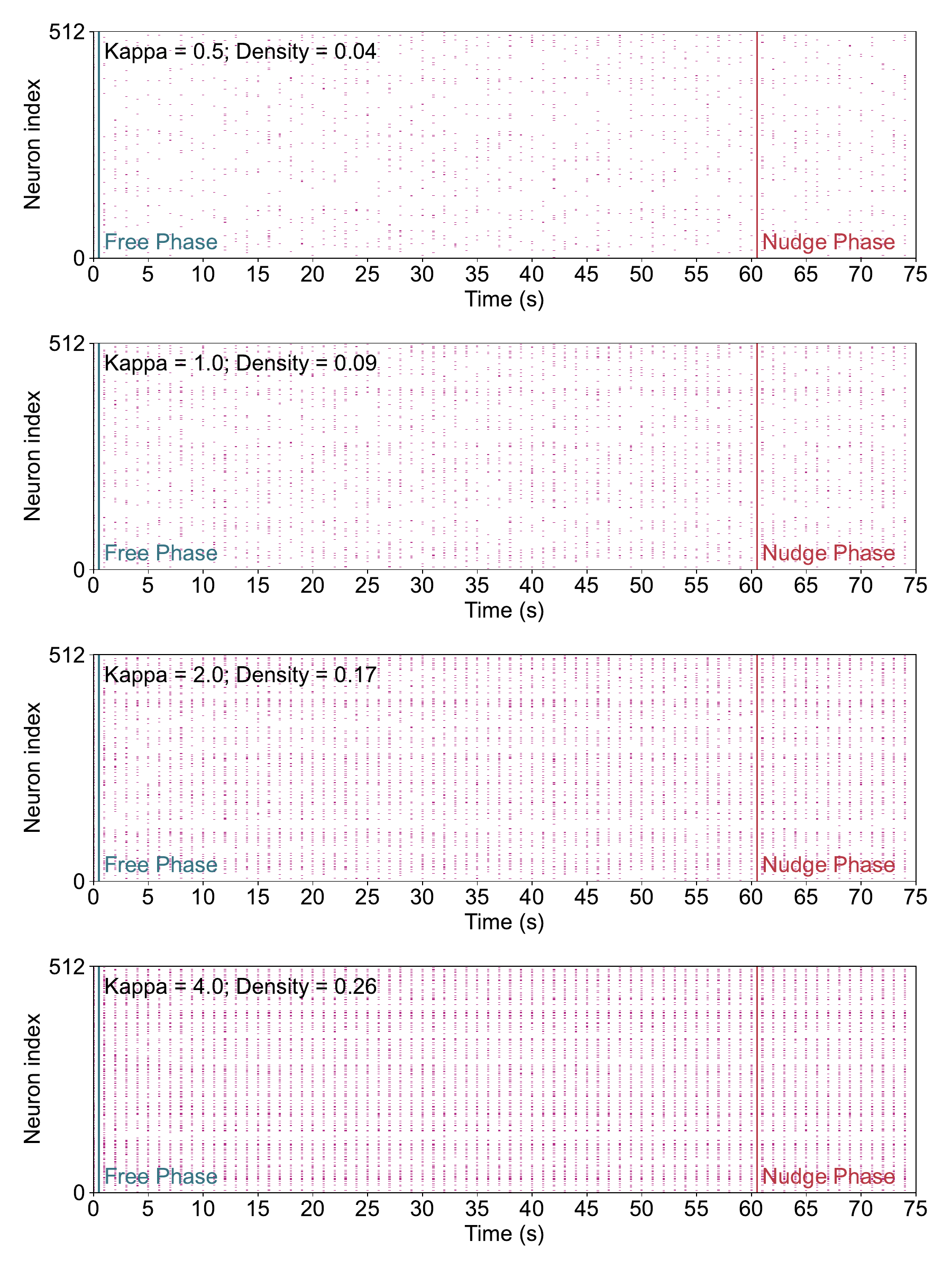}
  \caption{Spike raster plots of hidden-layer neurons trained on MNIST for different scaling factors $\kappa$. As $\kappa$ increases from 0.5 to 4.0, the firing density rises from 0.04 to 0.26, indicating more frequent spiking activity. Each plot shows neuron firing patterns during the free and nudged phases.
 }
  \label{fig:scalingfac}
\end{figure}

In addition, we analyze how the scaling factor $\kappa$ (Equation~\ref{eq:sigma}) influences network sparsity and the noise level. If firing becomes too low, error signals fail to propagate and learning stalls. In this regime, we increase $\kappa$ to raise spike rates. Conversely, if firing becomes too high, the phase contrast that drives EP weight updates is diminished. We then decrease $\kappa$ to restore a useful contrastive signal. To visualize sparsity in SNNs, we varied $\kappa$ to control the spike firing frequency, which is sampled from a Bernoulli distribution based on the membrane potentials. The results are demonstrated in Figure~\ref{fig:scalingfac}.

\section{Effects of Output Layer Inflation Factor}
\label{app:output_neuron}
\begin{figure}
  \centering
  \includegraphics[width=0.7\textwidth]{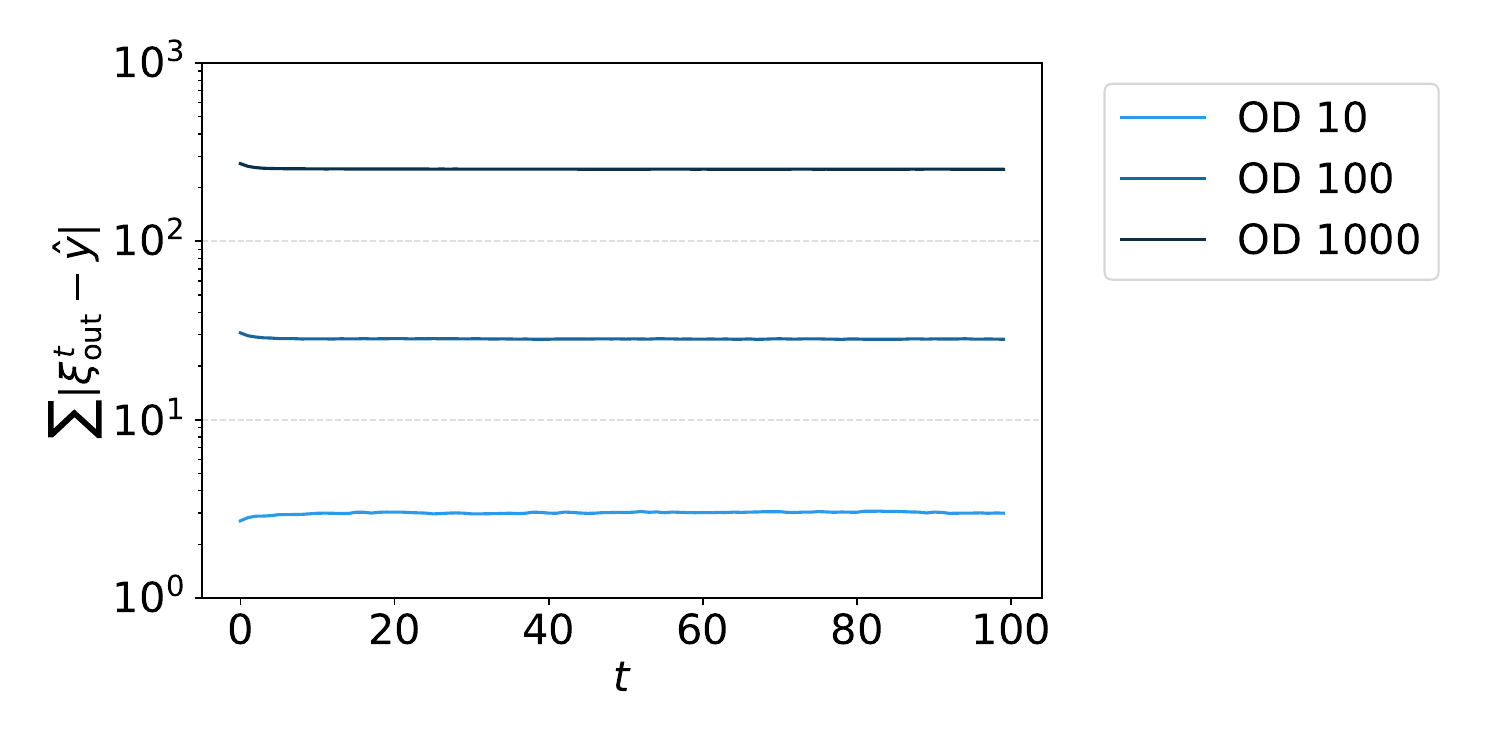}
  \caption{Summed magnitude of error signals at the output layer, averaged over 128 samples from the CIFAR-10 dataset using the 5C architecture. OD denotes the total number of output neurons (10, 100, and 1,000) corresponding to $N_{\rm perclass}=1, 10,$ and $100$, respectively. Increasing $N_{\rm perclass}$ amplifies the output-layer error signals, confirming stronger gradient propagation during training.}
  \label{app:fig:errsig}
\end{figure}

\begin{figure}
  \centering
  \includegraphics[width=0.7\textwidth]{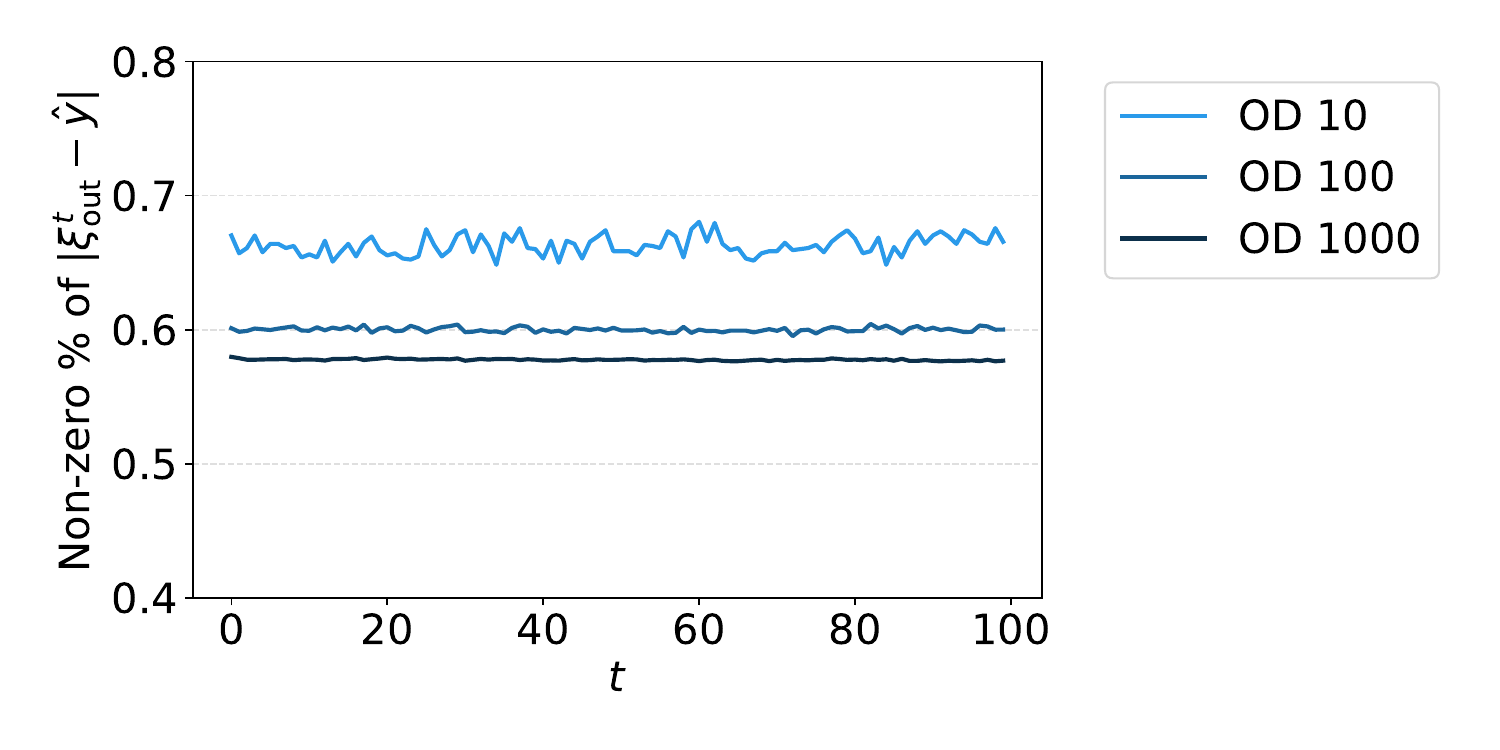}
  \caption{Percentage of non-zero error signals at the output layer, averaged over 128 samples from the CIFAR-10 dataset using the 5C architecture. OD denotes the total number of output neurons (10, 100, and 1,000), corresponding to $N_{\rm perclass}=1, 10,$ and $100$, respectively.}
  \label{app:fig:zerosoutput}
\end{figure}

In this section, we investigate the effect of the output-layer inflation factor $N_{\rm perclass}$ on the strength of error signals. We quantify how increasing the number of output neurons influences gradient propagation during training. The experiment is conducted using the 5C architecture on the CIFAR-10 dataset, where the error signals are evaluated by averaging over 128 random samples (Appendix~\ref{app:fig:errsig}). As $N_{\rm perclass}$ increases, a consistent rise in the summed error-signal magnitude is observed, indicating that output-layer inflation effectively amplifies gradient flow and enhances feedback propagation in the stochastic EP framework. Importantly, the percentage of non-zero error signals remains largely unchanged across different values of the inflation factor $N_{\rm perclass}$ (Figure~\ref{app:fig:zerosoutput}). However, since the total number of output neurons increases with larger $N_{\rm perclass}$, the overall magnitude of received error signals also grows accordingly.

\section{Proof of Equivalence for Stochastic EP Framework and Deterministic EP Framework under Mean-field Theory}
\label{app:expect_proof}
In this section, we provide the formal proof of Theorem~\ref{thm:stoeng}.

\begin{proof}
Assume throughout that (i) $\sigma(\cdot)=\rho(\cdot)$, where $\rho$ is the activation function used in the deterministic model, and (ii) conditioned on the membrane states $\xi$, the output spikes are independent Bernoulli random variables. By definition in Equation~\ref{eq:sto_engr}, the stochastic energy function is defined as:
\begin{equation}
E_{\rm Stoch}(x,\xi^t,w) =\tfrac{1}{2} \sum_{i=0}^{N_{\mathrm{t}}-1} {\|\xi^t_i\|}^2- \sum_{i=0}^{N_{\mathrm{t}}-2} \mathcal{B}(\sigma(\xi^t_{i}))^\mathsf{T}  w_i \mathcal{B}(\sigma(\xi^t_{i+1}))
\end{equation}
Taking expectation of $E_{\rm Stoch}(x,\xi^t,w)$:
\begin{equation}
\mathbb{E}\left[E_{\rm Stoch}(x,\xi^t,w)\right] = \tfrac{1}{2} \sum_{i=0}^{N_{\mathrm{t}}-1} {\|\xi^t_i\|}^2 - \sum_{i=0}^{N_{\mathrm{t}}-2} 
 \mathbb{E}\left[\mathcal{B}(\sigma(\xi^t_{i}))^\mathsf{T}  w_i \mathcal{B}(\sigma(\xi^t_{i+1}))\right]
\end{equation}
By the mean-field theory for Bernoulli distribution $\mathbb{E}[\mathcal{B}(\sigma(\xi^t))]=\sigma(\xi^t)=\rho(\xi^t)$ and independence across neurons,
\begin{align}
\mathbb{E}\left[\mathcal{B}(\sigma(\xi^t_i))\right] & =\rho(\xi^t_i) \\
\mathbb{E}\left[\mathcal{B}(\sigma(\xi^t_j))\mathcal{B}(\sigma(\xi^t_k))\right]
&=\rho(\xi^t_j)\rho(\xi^t_k)\quad(j\neq k)\\
 \mathbb{E}\left[\mathcal{B}(\sigma(\xi^t_{i}))^\mathsf{T}  w_i \mathcal{B}(\sigma(\xi^t_{i+1}))\right]   &=\rho(\xi^t_{i})^\mathsf{T}  w_i \rho(\xi^t_{i+1})
\end{align}
Hence, by the linearity of expectation, we have:
\begin{equation}
    \mathbb{E}\left[E_{\rm Stoch}(x,\xi^t,w)\right] = \tfrac{1}{2} \sum_{i=0}^{N_{\mathrm{t}}-1} {\|\xi^t_i\|}^2 - \sum_{i=0}^{N_{\mathrm{t}}-2} 
 \rho(\xi^t_{i})^\mathsf{T}  w_i \rho(\xi^t_{i+1}) = E(x,\xi^t,w)
\end{equation}
which gives $\mathbb{E}[E_{\mathrm{stoch}}(x,\xi^t,w)] = E(x,\xi^t,w)$. Since $E(x,\xi^t,w)$ is established as a Lyapunov function for the deterministic state dynamics \citep{scellier2017equilibrium}, the expected stochastic energy serves as a Lyapunov function in expectation and is monotonically decreasing along trajectories, implying convergence of the neuron states.
\end{proof}
Note that this convergence analysis is not tied to a specific noise distribution. As long as stochastic spike generation can be estimated by its expected firing rate, this proof can be extended with minor changes by replacing sampled spikes with their mean-field expectations. Thus, the same convergence argument applies.

\end{document}